# Spherical Fused Silica Cells Filled with Pure Helium for NMR-Magnetometry

Andreas Maul[1], Peter Blümler[1]*, Werner Heil[1], Anna Nikiel[1†], Ernst Otten[1], Andreas Petrich[2], Thomas Schmidt[2]

1. Institute of Physics, Johannes Gutenberg University, Staudingerweg 7, 55128 Mainz, Germany
2. ifw Günter-Köhler-Institut für Fügetechnik und Werkstoffprüfung GmbH, Otto-Schott-Str.13, 07745 Jena, Germany

† new address: Generaal Bentickstraat 46, 5623GX Eindhoven, The Netherlands

*corresponding author: bluemler@uni-mainz.de

**Abstract:**

High magnetic fields (> 1 T) are measured by NMR magnetometers with un-rivaled precision if the precessing spin sample provides long coherence times. The longest coherence times are found in diluted $^3$He samples, which can be hyperpolarized for sufficient signal strength. In order to have minimal influence on the homogeneity and value of the measured magnetic field the optimal container for the $^3$He should be a perfect sphere.

A fused silica sphere with an inner diameter of 8 mm and an outer diameter of 12 mm was made from two hemispheres by diffusion bonding leaving only a small hole for cleaning and evacuation. This hole was closed in vacuum by a $CO_2$ laser and the inner volume was filled with a few mbars of $^3$He via wall permeation. NMR-measurements on such a sample had coherence times of 5 min.

While the hemispheres were produced with < 1 μm deviation from sphericity the bonding left a step of ca. 50 μm at maximum. The influence of such a mismatch, its orientation and materials in the direct vicinity of the sample are analyzed by FEM-simulations and discussed in view of coherence times and absolute fields.

# Introduction:

In a recent publication[1] we presented measurements of high magnetic fields (> 1 T) with very high precision ($\delta B/B < 10^{-12}$) by a hyperpolarized $^3$He-NMR-probe. Such high precision is in demand for many fundamental physics experiments which test the standard model at low energies with extreme precision[2, 3]. In case of measuring the magnetic flux density via the detection of the frequency of the free induction-signal, it can be stated that the longer the observation time the higher the precision.

Therefore, finding samples with long spin coherence times (characterized by the time constant $T_2^*$ for the exponential signal decay) is of paramount importance to achieve higher precision. The possible achievable accuracy of the frequency and thus magnetic field measurement can be estimated according to the Cramér-Rao Lower Bound (CRLB)[4] and is given by[1]

$$\frac{\delta B}{B} \geq \frac{\sqrt{12}}{\gamma \cdot SNR \cdot \sqrt{f_{BW}} \cdot (T_2^*)^{3/2}} \cdot \frac{1}{B} \quad . \qquad [1]$$

Besides the characteristic power law $\sim (T_2^*)^{-3/2}$, the measurement sensitivity depends on the signal-to-noise ratio *SNR* in the bandwidth, $f_{BW}$, of the acquisition duration $T \approx T_2^*$ and the gyromagnetic ratio $\gamma$.

Since the longest coherence times are found in diluted gases we used $^3$He-gas in almost spherical glass cells which gave $T_2^*$ in the range of about one minute (for cell diameters in regime of ca. 1 cm[1]). In order to overcome the small number of nuclear spins in such a diluted medium they need to be hyperpolarized far beyond the thermal Boltzmann distribution by optically pumping via metastability exchange (MEOP = Metastability Exchange Optical Pumping[5, 6]). In this work[1] it was also demonstrated that the shape of the sample container has a substantial influence on the coherence time due to a distortion of an otherwise homogeneous external field $\vec{B} = B_0\,\hat{z}$ by the magnetization of the container. A prominent exception is the case of a container observing perfect spherical symmetry. In that case the field remains homogenous inside $\left(\vec{B}_{in} = B_{in}\,\hat{z}\right)$, but changes by a factor

$$\frac{B_{in}}{B_0} = \frac{9\mu}{(2\mu+1)(\mu+2) - 2(R_{in}/R_{out})^3(\mu-1)^2} \xrightarrow{|\chi|=|\mu-1|\ll 1} 1 - \frac{2}{9}\left[1-(R_{in}/R_{out})^3\right]\chi^2 \quad, [2]$$

where μ is the permeability of the wall material, and $R_{in}$ and $R_{out}$ are the inner and the outer container radii[7]. The permeability of the media inside and outside is considered to be μ = 1.

For para- or diamagnetic wall material having small susceptibility ($|\chi| = |\mu - 1| \ll 1$) the expansion in powers of $\chi$ on the right of eq. [2] yields a relative field drop inside scaling with $\chi^2$ in lowest order; hence it is extremely small. The literature value[8] for fused silica, e. g., is $\chi \approx -10^{-5}$, causing a relative field drop of order $10^{-10}$.

If we consider a small and smooth residual inhomogeneity of the field in the region of the NMR-probe we may approximate therein the change of field strength over a distance $\Delta \vec{r}$ by

$$\Delta B = \Delta \vec{r} \cdot \vec{\nabla} B \approx \Delta \vec{r} \cdot \vec{\nabla} B_z \quad [3]$$

with a constant field gradient $\vec{\nabla} B_z$. The Larmor-precession of an ensemble of immobile spins in a probe with radius $R$ would then get dephased relative to each other by the field gradient. In case of a gaseous probe, however, this gradient relaxation may be greatly suppressed by fast diffusion, known as motional narrowing. The effect may be explained and roughly estimated as follows: A single, typical random walk ($n$) starting from a point $P_n$ to another point $P_{n+1}$ on the inner probe surface will bridge in the average a distance of order $R$ and hence take a time of order $\Delta t_n = \mathcal{O}(R^2 / 2D)$ where $D$ is the diffusion constant. During this time the precession of the diffusing spin will accumulate a phase of

$$\begin{aligned}
\Delta \phi_n &= \gamma B_0 \Delta t_n + \mathcal{O}\left(\gamma \left(\vec{R}_{n+1} - \vec{R}_n\right) \vec{\nabla} B_z \Delta t_n\right) \\
&= \gamma B_0 \Delta t_n + \mathcal{O}\left(\frac{R^3 \gamma}{2D}\right)\left(\hat{d}_n \cdot \vec{\nabla} B_z\right) := \omega_0 \Delta t_n + \delta \phi_n
\end{aligned} \quad [4]$$

with $\hat{d}_n = \left(\vec{R}_{n+1} - \vec{R}_n\right) / \left|\vec{R}_{n+1} - \vec{R}_n\right|$. If the diffusion is fast enough, such that the phase spreading $\delta \phi_n$ occurring during such a random walk is much smaller than $2\pi$, than it takes a large number of them until the precession gets dephased, in particular since the $\delta \phi_n$ are uncorrelated in sign. Within a time interval $\tau = N \Delta t_n$ they sum up in the average to zero whereas their second moment grows with $N$:

$$\begin{aligned}
\left\langle \left(\delta \phi(t + \tau) - \delta \phi(t)\right)^2 \right\rangle_t &= \left(\sum_1^N \delta \phi_n\right)^2 = N \cdot \delta \phi_n^2 \\
&= \frac{\tau}{\Delta t_n} \cdot \delta \phi_n^2 = \tau \cdot \mathcal{O}\left(\frac{R^4 \gamma^2 \left|\vec{\nabla} B_z\right|^2}{2D}\right)
\end{aligned} \quad [5]$$

The coherence loss of an oscillation with fluctuating frequency within a time interval $\tau$ is defined by the decay of the corresponding autocorrelation function

$$K(\tau) = \left\langle \exp i\big(\phi(t+\tau) - \phi(t)\big) \right\rangle_t \, ,  \quad [6]$$

which is known to decay exponentially with half of the second moment of the phase fluctuation. Hence we get from eqs. [5] and [6]

$$K(\tau) = \exp\left[ i\omega_0\tau - \mathcal{O}\left(\frac{R^4\gamma^2 |\vec{\nabla} B_z|^2}{4D}\right)\tau \right] := \exp\left[\left(i\omega_0 - \frac{1}{T_{2\,(\text{est})}^{\text{dB}}}\right)\tau\right] \quad [7]$$

The structure of this estimate of the gradient caused relaxation rate, $1/T_{2\,(\text{est})}^{\text{dB}}$, is confirmed by the rigorous calculation in terms of diffusion modes[9]

$$\frac{1}{T_2^{\text{dB}}} = \frac{8R^4\gamma^2 |\vec{\nabla} B_z|^2}{175 D} \, . \quad [8]$$

The phase relaxation rate is further enhanced by the relaxation rate $1/T_1$ of the longitudinal polarisation, yielding the total transverse relaxation rate

$$\frac{1}{T_2^*} = \frac{1}{T_1} + \frac{1}{T_2^{\text{dB}}} \, . \quad [9]$$

Several mechanisms, like relaxation in field gradients and dipolar interactions, contribute to $1/T_1$[10]. However, for the conditions relevant for this work (small $R$, low $^3$He pressure, homogeneous field), longitudinal relaxation due to wall collisons[11] dominates all others. It is given by, $1/T_1^{\text{wall}} = \eta \cdot S/V = 3\eta/R$, where $\eta$ is the specific relaxivity of the wall material and $S/V$ is the surface/volume ratio. Hence, the coherence time is then well described in our case by

$$\frac{1}{T_2^*} \cong \frac{1}{T_1^{\text{wall}}} + \frac{1}{T_2^{\text{dB}}} = \frac{3\eta}{R} + \frac{8R^4\gamma^2 |\vec{\nabla} B_z|^2}{175 D} \quad [10]$$

with values for the self diffusion coefficient of $^3$He, given by $D = D_{0,\,^3\text{He}} \cdot (p_0/p) \cdot (T/273\,\text{K})^{1/2}$ [cm$^2$/s] ($D_{0,\,^3\text{He}} = 1880$ [cm$^2$/s] at $p_0 = 1$ mbar[12]) and $\gamma$, the gyromagnetic ratio of $^3$He, $\gamma_{^3He} = 2\pi \times 32.4340921(97)$ [MHz/T][13].

Equation [8] suggests a dramatic increase of $T_2^{\text{dB}}$ for decreasing sample size with the fourth power of its radius. However, in our earlier work this was only partially observed because all cells had a stem of fixed dimensions (see Fig. 1a), which introduced a strong deviation from sphericity and additionally the distortions became more pronounced for smaller sample sizes.

Besides reducing $T_2^*$ such geometric features also cause a strong dependence of $T_2^*$ on the sample orientation relative to the external field. In our example, longer $T_2^*$ was observed with the stem perpendicular to the field. Generating hollow spheres with high sphericity would therefore be very important to achieve longer $T_2^*$ and thus even higher sensitivities without dependence on orientation.

In earlier work[13] metal spheres were coated with glass and the metal later etched away by chemicals. Alternatively hollow glass spheres can be produced in drop-tower furnaces[14], however only with very thin walls, which will become problematic to contain the helium even at room temperature. Therefore, the goal of this work was to build thick walled fused silica spheres of high sphericity filled with a few mbar of pure $^3$He.

## Manufacturing of the Spheres:

In order to obtain closed fused silica cells with wall thicknesses in the millimeter range and high sphericity the following steps were taken:

a) Production and diffusion bonding of perfect fused silica hemispheres,
b) Evacuation trough a small hole and sealing it with a high power laser,
c) Filling with a defined helium pressure via wall permeation at elevated temperatures[15].

The individual steps will now be described in detail.

### a) diffusion bonding of fused silica hemispheres

Fused silica hemispheres were produced by the company Optrovision (Munich, Germany) which is specialized for the fabrication of high precision optical components. The hemispheres had an inner diameter of 8 mm and a wall thickness of 2 mm (this was a suggestion of the manufacturer to guarantee high quality) and were pitch polished[1] to reach a deviation from sphericity better than 1 μm. The quality of the inner and outer surface of the hemispheres were checked by interferometry relative to a negative specimen. Independent tests via a scanning interferometer (Luphos GmbH, Mainz, Germany) of the outer surface of the finished hemisphere verified a deviation from sphericity in order of 100 nm.

Two of these hemispheres were then attached via diffusion bonding (Hellma, Müllheim, Germany) to form an almost perfect hollow sphere. As a mechanical precision limit for the alignment of the hemispheres we (unfortunately, see below) didn't ask for better than 100 µm. In order to allow pressure equilibration during diffusion bonding, subsequent cleaning of the inner volume and for later evacuation a small rectangular hole of about $0.2 \times 0.1$ mm$^2$ in size was fabricated at the bonding equator. The latter production steps were also performed by Hellma Analytics.

---

[1] The spheres were polished to grade P3 (ISO10110) which specifies the maximum permissible number of micro-defects with size smaller than 1 μm.

The quality of the alignment of the hemispheres was determined via digital indicators and a confocal microscope (µsurf, Nanofocus AG, Oberhausen, Germany) to have a maximal mismatch of about ±50 µm.

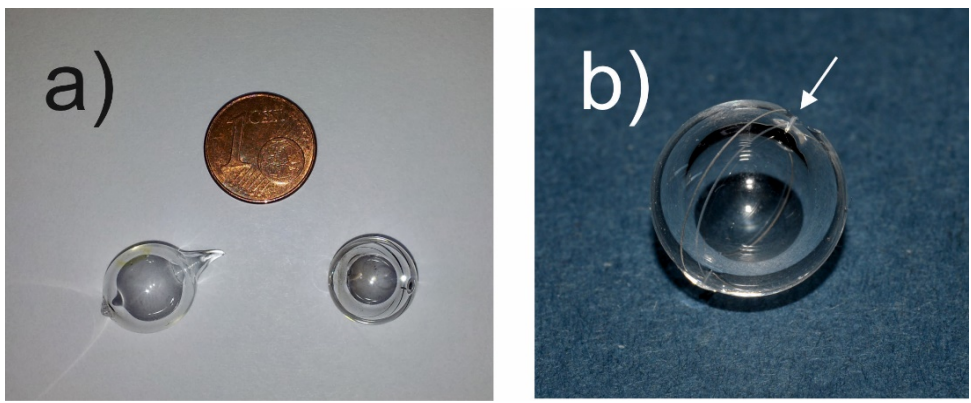

Fig. 1: a) Photographs of an old (left) cell with a stem and the new cell (right) below a Euro-cent for scale. b) close up of the new cell. The bright rim is the equator where the two hemispheres were bonded together. The sealed hole is indicated by the arrow. (color version online)

*b) evacuation and sealing with a high power laser*

Because the final sample has to contain helium in high purity it is necessary to evacuate and bake out the sphere prior to sealing and filling. The chosen procedure to achieve this was sealing the tiny hole inside a vacuum chamber by focusing a high power laser beam through an optical ZnSe-window which in difference to fused silica is transparent[2] at the wavelength (10.6 µm) of the used laser light (see Fig. 2).

---

[2] The IR edge of quartz falls between 4.5 and 5.0 µm. Therefore quartz is opaque for $CO_2$ laser light at a wavelength ~ 10.6 µm.

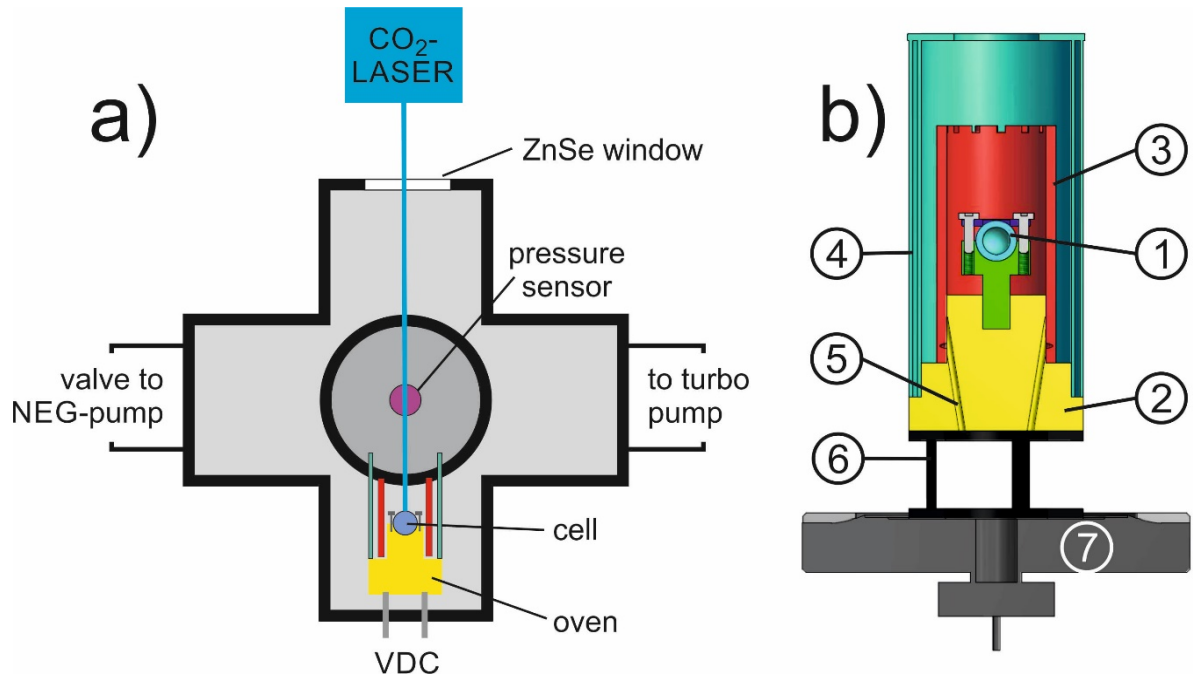

Fig. 2: a) Schematic drawing of the sealing setup. The gray "cross" in the center is a sketch of the vacuum chamber with 6 cylindrical extensions. b) close up of the oven/cell support part. 1) spherical fused silica cell (light blue) fixated to a Macor support (green) by a fused silica ring (dark blue) using screws (gray), 2) Macor support (yellow), 3) Macor cylinder (red) on which the heating wire is wrapped, 4) outer heat shield (turquoise), 5) wire feedthroughs, 6) stainless steel spacers (black), 7) CF flange (dark gray). (color version online)

Outbaking of the sphere was performed inside a home-build oven placed in the center of the vacuum chamber. This oven consisted of a ceramic support (Macor machinable ceramic) which encases the sphere from one side (bottom) while the top of the sphere is fixed against the Macor via a small fused silica ring (10 mm opening, fixed with 3 mm stainless steel screws) to match thermal expansions. The ceramic support was covered by a Macor cylinder with 2 mm notches to hold 12 windings of a resistive wire (0.5 mm thick Tantalum wire). The oven was isolated from the mounting flange via stainless steel spacers. The contacting wires were tripled in order to reduce the amount of heat inside the Macor feedthroughs. Applying 75 W (about 10-13 VDC and ca. 6 A) of electrical power gave a temperature of about 500 °C at the spot of the fused silica sphere (higher temperatures were avoided because Macor has a maximum temperature of 800 °C for continuous use). The sphere was baked out for about 10 h inside this oven while being evacuated by a turbo pump (Turbovac 50 from Leybold, Germany) with a getter (SORB-AC from saes getters, Italy) to about $10^{-5} – 10^{-6}$ Pa.

The laser system used for sealing the quartz sphere was a 200 W $CO_2$-Laser (Synrad Inc., Mukilteo, WA USA) located at ifw Günter-Köhler-institute for joining technics and material testing GmbH in Jena. The laser was carefully focused and adjusted via a scanner (Laser

Design GmbH, Essen, Germany) onto a small quartz plate which was temporarily placed on top of the sample. In a number of test experiments at Jena, optimal sealing conditions were found for a beam diameter of about 5 mm, 70 W laser power (86% power drop, ca. 300 W/cm$^2$ average power density at the sealing spot) and a pulse length of 15 s. The heating of the quartz happens solely via absorption within the first µm below the surface and subsequent heat conduction with the rather small conductivity of 1.38 Wm$^{-1}$K$^{-1}$. The latter set the time scale of the laser pulse in order to melt the silica down to a sufficient depth. A reduced beam diameter would reduce the amount of molten material hence risking a non-sealed cell. To increase the amount of melt right in the rectangular hole, a small quartz plug (2 mm long, 0.5 mm diameter) was inserted into it before sealing. This provision improved markedly the quality of the seal. The volume of the sphere could be evacuated through the partially plugged hole with a time constant of a few seconds (by a vacuum conductance reduced to $C \approx 10^{-4}$ l/s[16]). After irradiation with the laser the hole is not sealed throughout its full depth but only to a depth of ca. 1 mm. Three cells have been sealed this way.

Due to evaporation of silica, the laser-sealing left a shallow depletion (ca. 4-5 mm in diameter) on the outer surface and reached a central depth up to 100 µm as determined by scans with the confocal microscope. (A possible influence of this deviation from a perfect sphere on the magnetic field gradient inside will be discussed later.) The evaporated mass was about 0.5 mg as determined by weighting. At the sealing temperature, SiO$_2$ evaporates predominantly via dissociation into SiO + ½ O$_2$. Indeed the gas pressure in the vacuum chamber raised up to 10$^{-3}$ Pa during sealing, in spite of the large pumping power. In vacuum the evaporation rate may be estimated by the Hertz-Knudsen law $R \left[ \text{kg m}^{-2} \text{ s}^{-1} \right] = p_{\text{sat}}(T) / \sqrt{2\pi kT/m}$, which is the ratio of the saturation pressure over the rms-value of the transverse thermal velocity. $p_{\text{sat}}(T)$ may be read from Fig. 11 in review[17]. A depletion of 100 µm would then occur at a temperature close to 2100 K, lasting for 10 s and causing a local saturation pressure of about 20 Pa. These are reasonable numbers.

We were concerned that the temporary high partial O$_2$ pressure might leak by parts into the cell and be captured there. This would seriously degrade later on the MEOP-process in the $^3$He-discharge. However, we didn't see clear signs of molecular bands contaminating the He-spectrum, and optical pumping worked fine. Anyway, in a future approach we would like to avoid this risk by sealing the hole with a glass solder which melts at lower temperature and fits to fused silica.

*c) filling the cell with a defined helium pressure*

Fused silica is essentially impermeable to most gases, but helium, hydrogen and neon diffuse rather easily through this glass. In solids, the rate of diffusion increases steeply at higher temperatures and is proportional to the differential pressure. The selective diffusion of helium through fused silica is the basis for a method of purifying helium by essentially "screening out" contaminants by passing the gas through thin-walled fused silica tubes[15]. Because the permeation rate of helium is quite sensitive of the exact material composition, used helium isotope, temperature and geometry, we decided to extrapolate a rough time constant, $T_p$, for the permeation from[18] and give the process sufficient time (ca. 5 $T_p$) to reach equilibrium with the outside pressure. The time dependence for the permeation of $^4$He through fused silica was determined in[18] using

$$p(t) \approx p(0)\left[1 - \exp\left(-\frac{t}{T_p}\right)\right] \quad \text{with} \quad T_p = \frac{V\,d}{K\,S}, \qquad [11]$$

where $p$ is the pressure, $K$ is a permeation velocity for a certain material and temperature and $d$ is the wall thickness. For $^4$He diffusion through fused silica at temperature $T = 500°C$ the permeation velocity was determined to be $K = (4 \pm 0.5) \cdot 10^{-11}$ m$^2$/s[18]. We used this number also for $^3$He as a conservative lower limit, since the lighter isotope should diffuse somewhat faster. That results in a characteristic permeation time constant of $T_p \approx 1$ day for our spherical cells with $Vd/S = 2.67$ mm$^2$. The filling temperature of 500 °C was chosen as a compromise between materials stability, undemanding heating and reasonably fast filling rate, because eq. [11] gives a conservative $T_p$ of about 1 day for these conditions.

To fill the spheres we used the vacuum chamber depicted in Fig. 1. After evacuation the valve to the turbo pump was closed and the chamber filled with the destined pressure of $^3$He (here 1.5 mbar). The oven was then set to a temperature of about 500 °C and left for 5 days to assure equilibrium. During this time the NEG-pump was active in order to catch H$_2$ which might emanate from the chamber walls and permeate into the cell. After cooling the sample down, the pressure of the outer reservoir was checked again to assure for the absence of leakage.

Because the experimental protocol involves the ignition of an electrical discharge inside the sphere for MEOP[5], the effortlessness to ignite a plasma is a crude indicator on the amount of $^3$He inside the cell while its color (deep purple) can be used as a rough test on its purity.

## Experimental and Discussion

To test for a possible increase of the transverse relaxation times in the improved fused silica cells, we repeated the experiments at a clinical 1.5 T whole-body MRI-tomograph, as described in[1]. There $T_2^*$ was shortened by the inhomogeneous magnetization of the non-spherical container. This was clearly demonstrated there by the dependence of $T_2^*$ on the reorientation of the sample relative to the magnetic field and the results agreed fairly well with simulations. For a cell with a pronounced stem (cf. Fig. 1a) and an outer diameter of 15 mm (inner diameter 13 mm) a $T_2^*$ of 67 s was measured when filled with 1 mbar of $^3$He and the stem oriented perpendicular to the main field of 1.5 T , while $T_2^*$ reduced below 9 s for the parallel orientation.

When using one of the glass cells produced as described in this publication, the measured characteristic $T_2^*$ times increased to values between 3 to 5 min (see Fig. 3b). The *SNR* was about 300 related to a bandwidth of 1 Hz. The decrease in *SNR* by a factor of ~ 6 relative to previously determined values on the 15 mm cells[1] is due to the smaller volume of the sample and, consequently, a smaller filling factor of the same surface NMR-coil as already used in[1]. Of course the *SNR* could be improved by using another NMR-coil, which should be better adopted to the sample size and shape, however risking a disturbance of the field homogeneity when placing metal wire close to the cell. If doing so the use of zero-susceptibility wire is advisable[19]. Additionally the *SNR* can be pushed by using higher $^3$He pressures (possibly up to some 100 mbar[20]), however on the cost of $T_2^*$.

The determined $T_2^*$ is far below the theoretically expected value of $T_2^{dB} \approx 2.7$ h estimated from the measured magnet's homogeneity of about $5 \cdot 10^{-6}$ T/m using eq. [10]. The major part of this drop is due to enhanced longitudinal wall-relaxation, which is more pronounced for smaller cells and will at some point limit the gain in $T_2^{dB}$ by shrinking the container size. $T_1$ was directly measured by monitoring the decrease of the signal amplitude in a series of delayed excitations with a small flip angle. Observing a $T_1 \cong T_1^{wall} = 24$ min in the $R_{in} = 4$ mm cell results in $\eta = 9.3 \cdot 10^{-7}$ m/s for our cells. This will then also define the upper limit for $T_2^* \leq$ 24 min. However, this still doesn't explain the full drop and the variability of the measured $T_2^*$.

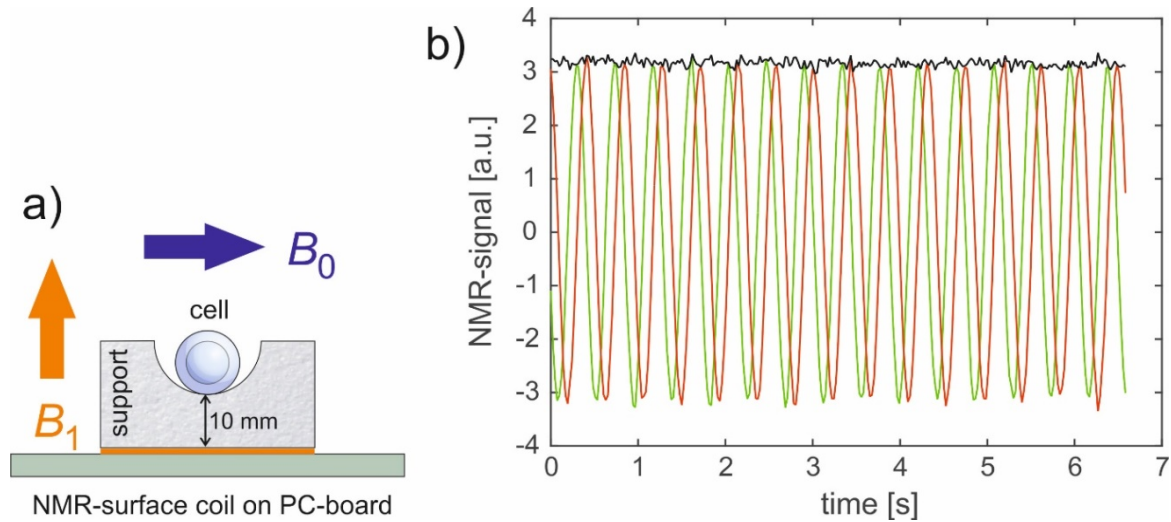

Fig. 3:   a) Schematic representation of the measurement geometry including the static magnetic field, $B_0$, and the RF-field, $B_1$, for excitation and detection. For a detailed description, see[1]. b) Detected NMR-signal with a dwell time of 24.2 ms. For technical reasons the total measurement time was limited to ca. 6 s. Shown are the real (red) and the imaginary (green) part of the signal. The black envelope is its magnitude from which a coherence time of $T_2^* = 340$ s $\pm$ 73 s was calculated for this particular run. (color version online)

To get a qualitative explanation for this observation, FEM-simulations (Comsol Multiphysics 5.1) were carried out placing two hemispheres onto each other but with a tiny offset, $\Delta r$, of their symmetry axis (measured to be $\Delta r \leq 50$ μm, see above) and exposing them to perfectly homogeneous magnetic flux of 1.5 T. However, determining the influence of such a small geometric mismatch on the magnetic field inside a much bigger object requires high precision. For this reason, full 3D simulations of the setup were beyond the computing power of our PC system. Therefore, the simulations were done in "pseudo-3D" axisymmetric mode (cf. Fig. 4a), which allowed testing with the necessary accuracy. However, this implied that the direction of the magnetic field had to be chosen parallel to the axis and normal to the equatorial plane with the small step. This geometric mismatch, $\Delta r$, had to be emulated by placing a slightly smaller (or bigger) hemisphere on top of the other rather than just shifting two identical ones in one direction only. This will give a worse estimate of the real effect, because the step then spans the entire circumference of the sphere and generates a much stronger deterioration of the magnetic field. The same argument holds for the wrong orientation of the magnetic field relative to the Styrofoam support (cf. Fig. 3a and Fig. 4a), again causing a much stronger effect on the field homogeneity in this orientation. Therefore, the reported values can be understood as a worse-case scenario.

Besides the geometry the used magnetic volume susceptibilities (all at 25°C in SI units) are of crucial importance for the result: $\chi_{\text{Fused Silica}}{}^8 = -1.103 \cdot 10^{-5}$, $\chi_{\text{Air}}{}^{21} = +3.7 \cdot 10^{-7}$, and $\chi_{\text{Polymer}} \approx 0$ for the Styrofoam support below the sphere. The latter cannot be neglected, because the used XPS-Styrofoam is a polymer foam with closed cells filled with a vapor that can be different from air (depending on the production process), hence replacing the slightly paramagnetic air (cf. Fig. 4b). Since we do not know the origin of the used Styrofoam we estimated its influence in a worst-case scenario as being filled with vacuum. The very dilute $^3$He gas inside the sphere was also approximated by vacuum ($\chi = 0$). The influence of the copper of the NMR-surface coil was also tested and found to be negligible and omitted in the other simulations. The same holds for the geometric flaw caused by sealing the cell with the laser. Its deviation from sphericity is in the order of the mismatch between the hemispheres but much smoother and due to the 2 mm thick fused silica walls too far away from the inner volume to produce a significant contribution to the internal gradients (this can be appreciated from Fig. 4a/b).

Simulating and averaging the internal gradients over the inner volume for non-shifted hemispheres (i.e. a perfect sphere or $\Delta r = 0$) gave residual values in the order of $\left|\vec{\nabla} B_z\right| = 10^{-8}$ T/m which are induced and depend very much on the chosen mesh size. But they are significantly lower than the external gradient of the magnet. Table 1 compares the simulated gradients $\left|\vec{\nabla} B_z\right|$ for different displacements, $\Delta r$, of the hemispheres with the equator normal to the direction to the external field (cf. Fig. 4a). From these values $T_2^*$ was calculated according to eq. [10]. It is important to note that eq. [10] can be only used to approximate the influence of such a localized field gradients on the coherence time, because eq. [10] was derived for a spatially homogeneous gradient$^9$. In our case a more realistic result would demand a Monte-Carlo simulation similar to previous work as described in$^{22, 23}$.

The maximal size of the mismatch was measured to be in the order of 50 µm for our sample. From the simulated values in Tab. 1 a $\Delta r$ of about 20 µm would best explain the observed $T_2^*$ of ca. 3 min. As it was mentioned above the influence of the mismatch is overestimated in the 2D axisymmetric simulation and a value of about 20 µm seems to be very reasonable for its averaged size over the entire perimeter. The homogeneity of the internal field is further deteriorated when the symmetry of the environment is disturbed by replacing the paramagnetic air by the Styrofoam support. This causes an almost linear gradient over the inner cell volume as depicted in Fig. 4b, whose value can be estimated from Tab. 1 to be about $10^{-5}$ T/m. Taken together these simulations give a good explanation for the observed $T_2^*$ values however they do not yet explain their variability in different runs.

We are convinced that this latter effect can be explained by the angular position of the equatorial plane relative to the magnetic field which was not explicitly controlled in the experiment, since we discovered the mismatch only thereafter. As already shown in[1] for a much bigger flaw, the position of such a step-like feature can easily change $|\vec{\nabla} B_z|$ by a factor of two when going from parallel to perpendicular orientation relative to the external field. Taken together, these simulations can explain the observed $T_2^*$ values and their variability (values between 3 to 5 min.

Table 1: 2D axisymmetric simulation of the internal gradients, $|\vec{\nabla} B_z|$, in a sample of two fused silica-hemispheres with different radii (bottom hemisphere with $R$ = 4 mm, top hemisphere with $R$ = 4 mm + $\Delta r$). Their equator is aligned normal to the external flux of 1.5 T and the resulting $T_2^*$ was calculated according to eq. [10] with $\eta = 9.3 \cdot 10^{-7}$ m/s as the surface relaxivity (setting an upper limit of 24 min to $T_2^*$). The magnitude of all gradient components ($\partial B_z / \partial i$ with $i = r, z$) were averaged over the inner volume for cells with and without a Styrofoam support (see Fig. 3a). Figure 4 shows some screenshots from these simulations. The last column shows the change of the average magnetic field, $\Delta B(\Delta r)$, inside the sphere relative to the external field (cf. eq. [12]) and to that of a perfect sphere, $\Delta B(0)$. See text for details.

| | without support | | with support | | without support |
|---|---|---|---|---|---|
| $\Delta r$ [μm] | $\|\vec{\nabla} B_z\|$ [T/m] | $T_2^*$ [min] | $\|\vec{\nabla} B_z\|$ [T/m] | $T_2^*$ [min] | $\Delta B(\Delta r) \cdot 10^9$ |
| 0 | $7.0 \cdot 10^{-9}$ | 23.93 | $8.55 \cdot 10^{-6}$ | 16.91 | 0 |
| 5 | $1.16 \cdot 10^{-5}$ | 13.66 | $1.18 \cdot 10^{-5}$ | 13.47 | 0.01 |
| 10 | $2.13 \cdot 10^{-5}$ | 6.88 | $2.68 \cdot 10^{-5}$ | 4.88 | 0.05 |
| 20 | $3.88 \cdot 10^{-5}$ | 2.62 | $4.27 \cdot 10^{-5}$ | 2.21 | 0.16 |
| 50 | $8.27 \cdot 10^{-5}$ | 0.64 | $8.39 \cdot 10^{-5}$ | 0.61 | 0.82 |
| 100 | $1.39 \cdot 10^{-4}$ | 0.23 | $1.42 \cdot 10^{-4}$ | 0.22 | 2.78 |
| 200 | $2.34 \cdot 10^{-4}$ | 0.08 | $2.35 \cdot 10^{-4}$ | 0.08 | 8.94 |
| 500 | $4.25 \cdot 10^{-4}$ | 0.02 | $4.26 \cdot 10^{-4}$ | 0.02 | 38.00 |

Originally our attempt was to measure relative changes in strong magnetic fields with high precision. Based on these results we had several requests if this instrument could also be used for absolute field measurements (e.g. for the g-2 experiment[3]). Owing to this we also simulated the influence of the slight mismatch, $\Delta r$, of the two hemispheres on the change of internal magnetic flux relative to the preset external field, $B_0$, as defined as

$$\Delta B(\Delta r) := \frac{B_0 - \langle B_z(\Delta r) \rangle_{in}}{B_0} \quad , \qquad [12]$$

where $\langle B_z(\Delta r) \rangle_{in}$ is the flux averaged over the inner volume of the cell. Table 1 shows that the values of $\Delta B(\Delta r)$ are in the range of $10^{-11}$ to $10^{-9}$ (again simulated in 2D axisymmetric geometry of two hemispheres with radii of 4 mm differing by $\Delta r$). Additionally, one should note that $\Delta B(\Delta r)$ in Tab. 1 was calculated relative to that of a perfect sphere ($\Delta r = 0$). The flux in a perfect hollow sphere with vacuum in- and outside is given by eq. [2] and amounts to $\Delta B^{vac}(0) = 1.9 \cdot 10^{-11}$. In our simulations we could reproduce this value to $\Delta B^{vac}(0) = 1.5 \cdot 10^{-11}$ due to numerical inaccuracies. One should be aware that this value changes dramatically to $\Delta B^{air}(0) = -1.2335 \cdot 10^{-7}$ when replacing the outer vacuum by the slightly paramagnetic air.

Table 1 shows that the magnetic fields inside cells with $\Delta r$ in the range of 100 μm should only deviate in the order of $10^{-10}$ relative to $B_0$. This effect is probably even smaller, if two identical hemispheres could be simulated in 3D being only shifted by $\Delta r$ (rather than differing in their radii), because the relative field deteriorations on opposite walls should cancel as suggested by Fig. 4d, since every inward feature is paired by an outward feature of the same size and causing very similar field changes of opposite sign, which will cancel each other. Hence the values of $\Delta B$ in Table 1 are too pessimistic.

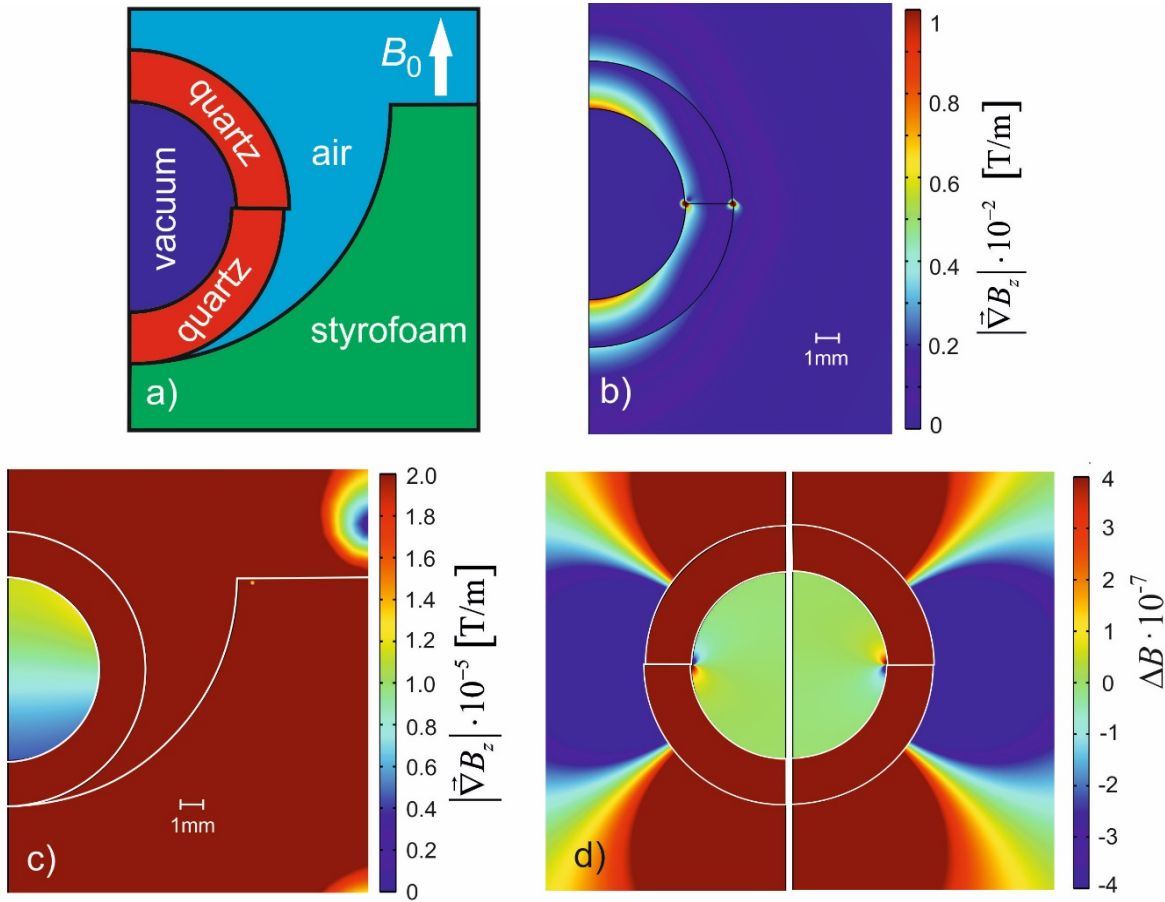

Fig. 4: 2D axisymmetric FEM-simulations of two hemispheres: a) Sketch of the geometry and illustrating the materials. Step size not to scale. Note the wrong direction of $B_0$ (cf. Fig. 3a). b) $|\vec{\nabla} B_z|$ of the hemispheres in air with $\Delta r = 50$ μm. Colors marking a range from 0 to $2 \cdot 10^{-2}$ T/m. c) same as b) but now as a perfect sphere ($\Delta r = 0$) with a lower support of Styrofoam (colors from 0 to $2 \cdot 10^{-5}$ T/m). d) $\Delta B$ (cf. eq. [12]) for hemispheres with radii differing by $\Delta r = 50$ μm in vacuum. In the right graph the bigger ($R = 4$ mm $+ \Delta r$) hemisphere is on top, while it is at the bottom on the left. The colors stretch from $-4 \cdot 10^{-7}$ to $+4 \cdot 10^{-7}$. Similar field deviations at the outside of the step are covered by the expected[7] exterior dipolar field distribution. Please note that values exceeding the color range are displays by the colors of the limiting values. (color version online)

## Conclusion and Outlook:

We demonstrated that thick walled, hollow fused silica spheres with diameters in the order of a centimeter can be produced with deviation from sphericity in the 50 μm range. We also managed to seal these spheres in vacuum using a $CO_2$ laser and subsequently filled them with a few mbars of $^3$He via wall permeation at elevated temperature. Overall the deviations from sphericity were well below 1 μm except for the equatorial plane where the two hemispheres were bonded with a tiny mismatch leaving a small step of maximum ~ 50 μm. With such cells the free induction decay of the precessing $^3$He sample spins were measured in the homogeneous magnetic field $B = 1.5$ T of an NMR scanner resulting in characteristic $T_2^*$ times of up to 5 min. We showed that the transverse relaxation rate $1/T_2^*$ is still dominated by inner gradients due to the tiny mismatch in bonding the hemispheres. The producers of the spheres however claim that they can principally produce spheres with better alignments on demand which can be important for attempts to measure absolute magnetic fields by this method (e.g. for the g-2 muon experiment). We plan to repeat the experiments with such improved cells to explore the full potential of this technique for magnetometry.

For this reason we also presented simulations to emphasize the importance of the environment for such high precision experiments. These simulations were aimed to be worst case scenarios to allow for a conservative planning of future experiments.

## Acknowledgement:

The authors want to thank M. Terekhov for assistance in acquiring the NMR-signal on the clinical 1.5 whole body scanner, which was gratefully permitted by L. M. Schreiber (both University clinics Mainz). A. Best and K. Koynov (MPI-Polymerforschung, Mainz) have to be acknowledged for performing the confocal microscopy. The German Research Foundation (grant no. HE 2308/16-1) supported this work financially.